\documentclass{article}

\PassOptionsToPackage{numbers,sort&compress}{natbib}

\usepackage[final,main]{neurips_2026}

\usepackage[utf8]{inputenc}
\usepackage[T1]{fontenc}

\usepackage{graphicx}
\usepackage{soul}
\setuldepth{foobar}

\usepackage{tcolorbox}
\usepackage{enumitem}
\usepackage{array}
\usepackage{amsmath}

\usepackage{amssymb}

\usepackage{amsthm}

\newtheoremstyle{tight-italic}%
  {5pt}%   Space above
  {5pt}%   Space below
  {\itshape}% Body font (italic)
  {}%     Indent amount
  {\bfseries}% Head font
  {.}%    Punctuation after head
  { }%    Space after head
  {}%     Head spec

\theoremstyle{tight-italic}

\AtBeginEnvironment{definition}{\pushQED{\qed}}
\AtEndEnvironment{definition}{\popQED}

\usepackage{mathrsfs}
\usepackage{tikz-cd}
\usetikzlibrary{decorations.pathmorphing}
\usepackage{xcolor, colortbl}
\usepackage{booktabs}
\usepackage{multirow}
\usepackage{makecell}
\usepackage{pifont}
\usepackage[usestackEOL]{stackengine}
\usepackage{wrapfig}

\usepackage{url}
\usepackage{nicefrac}
\usepackage{microtype}

\usepackage{hyperref}
\usepackage[capitalize]{cleveref}

\title{A Steerable Deep Network for Model-Free\\Diffusion MRI Registration}

\author{%
  Gianfranco Cort\'{e}s\textsuperscript{1} \\
  Department of CISE \\
  University of Florida
  \And
  Xiaoda Qu \\
  Department of Statistics \\
  University of Florida
  \And
  Baba C. Vemuri \\
  Department of CISE \\
  University of Florida
}

\begin{document}
\maketitle
\footnotetext[1]{Correspondence: \texttt{gcortes@ufl.edu}.}
\setcounter{footnote}{1}

\begin{abstract}
    % Nonrigid registration is vital to medical image analysis but remains challenging for diffusion MRI (dMRI) due to its high-dimensional, orientation-dependent nature. We present a novel, geometric deep learning framework for {\it model-free}, nonrigid registration of raw dMRI data. We present a novel formulation of the dMRI registration problem as an equivariant diffeomorphism of position-and-orientation space. Central to our method is an $\mathrm{SE}(3)$-equivariant UNet that generates velocity fields while preserving the geometric properties of a raw dMRI's spatio-angular domain. We introduce a new loss function based on the maximum mean discrepancy in the Fourier space, implicitly matching ensemble average propagators across images being registered. Experimental results on HCP and OASIS clinical-grade dMRI data demonstrate competitive performance compared to state-of-the-art approaches, with the added advantage of bypassing the overhead for estimating derived representations. This work establishes a foundation for data-driven, geometry-aware dMRI registration directly in the acquisition space.
    
    Nonrigid registration is vital to medical image analysis but remains challenging for diffusion MRI (dMRI) due to its high-dimensional, spatio-angular dependence. We present a novel, geometric deep learning framework for {\it model-free}, nonrigid registration of raw dMRI data. We formulate registration in the native spatio-angular acquisition space under the joint action of 3D roto-translations, $\mathrm{SE}(3)$. Central to our framework is the equivalence between equivariant spatio-angular convolutions and 3D steerable convolutions on spherical-harmonic coefficient fields, which we exploit to construct an $\mathrm{SE}(3)$-equivariant U-Net. A by-product of this design choice is freedom from having to augment the data with roto-translated versions of itself. We also formulate a maximum mean discrepancy (MMD) loss between water-displacement distributions, evaluated in Fourier space as a weighted sum of squared signal differences, and extend it with a local contrast correction that reduces to a local correlation. In controlled comparisons, including a known deformation of real tissue, the raw signal yields markedly more accurate alignment than fractional anisotropy under the same optimizer. On held-out HCP subjects and on a held-out clinical-grade cohort (AOMIC) from a different scanner, our method improves anatomical overlap over ANTs SyN on fractional anisotropy, \texttt{mrregister} on fODFs, DDMReg and SynthMorph, without estimating any derived representation, and the steerable network matches a conventional U-Net with fewer parameters while being unaffected by grid rotations of its inputs. This work establishes a foundation for data-driven, geometry-aware dMRI registration directly in the acquisition space.
\end{abstract}
\section{Introduction}
\label{sec:intro}

Diffusion-weighted magnetic resonance imaging (dMRI) is a noninvasive imaging technique that probes tissue microstructure by sensitizing the MR signal to the direction-dependent diffusion of water molecules \cite{lebihan1992}. To interpret the measured signal, practitioners choose and fit a model of local water diffusion, yielding a \textit{derived representation}\footnote{We use this terminology to avoid confusion with generative diffusion models.} of the signal from which microstructural indices are extracted \cite{hagmann2006}. For example, the diffusion tensor model assumes that water diffusion is locally Gaussian, parameterized by a ($3 \times 3$) symmetric positive definite matrix at each spatial position~\cite{basser1994}. The resulting derived representation is a diffusion tensor image (DTI), from which features such as fractional anisotropy and the principal diffusion direction can be computed \cite{MCGRAW04}.

Comparing these features across time points or subjects requires that the dMRI data be brought into anatomical alignment, a process known as \textit{registration} \cite{sotiras2013}. The predominant approach to dMRI registration is to design algorithms around a chosen derived representation \cite{zhang2006,cheng2009,raffelt2011,zhang2021,bouza2023}, thereby tying the alignment to a particular modeling choice. In this paper, we formulate diffeomorphic registration directly on dMRI's acquisition domain, allowing the choice of derived representation for downstream analysis to be made independently from the registration task.

A dMRI signal can be modeled continuously by a function $S \colon \mathbb{PQ} \to \mathbb{R}$, where $\mathbb{PQ}$ is a space of pairs $(\mathbf{p}, \mathbf{q})$ composed of a spatial position $\mathbf{p} \in \mathbb{R}^3$ and a diffusion-encoding wavevector $\mathbf{q} \in \mathbb{R}^3$ \cite{callaghan1991}. In an acquisition, this signal is sampled as a collection of 3D images $S(\cdot, \mathbf{q}_k)$ indexed by finitely many wavevectors $\mathbf{q}_k$. 

\begin{figure*}[t]
\begingroup
\renewcommand{\arraystretch}{1.5}
\centering
\begin{tabular}{c!{\vrule width 1.5pt}c|c|c}
    \cellcolor[gray]{0.85} Input & \cellcolor[gray]{0.95} Scalar images & \cellcolor[gray]{0.95} Derived representations & \cellcolor[gray]{0.95} \\
    \cline{1-3}
    \cellcolor[gray]{0.85} Examples & sMRI, CT, FA & DT, fODF, GMM & \cellcolor[gray]{0.95}\multirow{-2}{*}{Raw dMRI signals} \\
    \hline
    \cellcolor[gray]{0.85}\Centerstack{Registration \\ schematic} &
    \begin{tikzcd}
    \mathbb{R}^3 \arrow[r, "I_{\mathrm{F}}"] \arrow[d, "\Phi", <->] & \mathbb{R} \\
    \mathbb{R}^3 \arrow[r, "I_{\mathrm{M}}"] & \mathbb{R}
\end{tikzcd} &
\begin{tikzcd}
    \mathbb{R}^3 \arrow[r, "S^{\mathscr{D}}_{\mathrm{F}}"] \arrow[d, "\varphi", <->] & \mathscr{D} \\
    \mathbb{R}^3 \arrow[r, "S^{\mathscr{D}}_{\mathrm{M}}"] & \mathscr{D} \arrow[u, squiggly, <->, "\textcolor{purple}{\text{reorient}}", draw=purple]
\end{tikzcd} &
\begin{tikzcd}
    \mathbb{PQ} \arrow[r, "S_{\mathrm{F}}"] \arrow[d, "\Phi", <->] & \mathbb{R} \\
    \mathbb{PQ} \arrow[r, "S_{\mathrm{M}}"] & \mathbb{R}
\end{tikzcd}
 \\
\end{tabular}
\caption{Does registration on $\mathbb{PQ}$ introduce additional degrees of freedom, or do its admissible alignments reduce to spatial deformations paired with local reorientation, as in derived-representation registration? Here, $\mathscr{D}$ denotes a derived representation's value space, e.g.\ $\mathscr{D}=\mathrm{SPD}(3)$ for DTI.}
\label{fig:reg-types}
\endgroup
\end{figure*}

Given two 3D images $I_\mathrm{M}, I_\mathrm{F}\colon \mathbb{R}^3 \to \mathbb{R}$, diffeomorphic registration seeks an orientation-preserving diffeomorphism $\Phi$ ($\det D\Phi > 0$) such that $I_\mathrm{M} \circ \Phi$ aligns with $I_\mathrm{F}$ \cite{beg2005}. Carrying this formulation over to dMRI signals by replacing $\mathbb{R}^3$ with $\mathbb{PQ}$ might suggest additional degrees of freedom beyond the spatial deformation and local reorientation typically used in derived-representation registration (\cref{fig:reg-types}). However, an unrestricted diffeomorphism of $\mathbb{PQ}$ would ignore the distinct roles of spatial positions ($\mathbb{P}$-space) and wavevectors ($\mathbb{Q}$-space) and permit physically implausible transformations. Hence, we characterize which alignments are admissible using the geometry of $\mathbb{PQ}$-space.

\begin{wrapfigure}{r}{0.36\textwidth}
  \centering
  \begin{tikzpicture}[x=1cm,y=1cm,font=\scriptsize,line width=0.35pt]
  \definecolor{shellouter}{RGB}{65,105,225}
  \colorlet{shellinner}{shellouter!55!white}
  % A light frame groups the base and fiber without competing with the diagram.
  \draw[black!25,line width=0.4pt,rounded corners=3pt]
    (-0.2,-0.55) rectangle (3.2,3.5);
  % A 2x2-cell lattice of spatial positions in perspective; p is the central point.
  \foreach \j in {0,...,2}
    \draw[black!15] ({0.4*\j},{0.3*\j}) -- ({2.2+0.4*\j},{0.3*\j});
  \foreach \i in {0,...,2}
    \draw[black!15] ({1.1*\i},0) -- ({1.1*\i+0.8},0.6);
  \foreach \j in {0,...,2}
    \foreach \i in {0,...,2}
      \fill[black!40] ({1.1*\i+0.4*\j},{0.3*\j}) circle (0.8pt);
  \node[anchor=north,font=\footnotesize,text=black!75,inner sep=1.5pt]
    at (1.5,-0.12) {Base: $\mathbb{P}$};
  \fill[black!75] (1.5,0.3) circle (1.5pt);
  \node[below right,inner sep=1.5pt,yshift=0.5pt] at (1.5,0.3) {$\mathbf{p}$};
  % Concentric shells in Q-space as shaded balls, one shade of blue per shell.
  \shade[ball color=shellouter!40!white] (1.5,2.13) circle (0.44);
  \shade[ball color=shellouter,opacity=0.45] (1.5,2.13) circle (0.84);
  \draw[shellouter!75!black,line width=0.3pt] (1.5,2.13) circle (0.84);
  % Draw over the shells so the guide remains visible up to the fiber origin.
  \draw[densely dashed,black!45] (1.5,0.3) -- (1.5,2.13);
  \foreach \a in {20,85,155,215,290,340}
    \filldraw[fill=shellouter!75!black,draw=white,line width=0.3pt] ({1.5+0.84*cos(\a)},{2.13+0.84*sin(\a)}) circle (1.2pt);
  \foreach \a in {45,130,195,265}
    \filldraw[fill=shellouter!75!black,draw=white,line width=0.3pt] ({1.5+0.44*cos(\a)},{2.13+0.44*sin(\a)}) circle (1.2pt);
  \node[anchor=south,font=\footnotesize,text=black!75,inner sep=1.5pt]
    at (1.5,3.05) {Fiber: $\mathbb{Q}$};
  \draw[black!45] (2.29,1.84) -- (2.65,1.7);
  \node[right,inner sep=0.8pt,yshift=-1.2pt] at (2.65,1.7) {$\mathbf{q}$};
\end{tikzpicture}
  \caption{We view $\mathbb{PQ}$-space as a vector bundle over $\mathbb{P}$-space with shell-sampled fibers.}
  \label{fig:pq-bundle}
\end{wrapfigure}

The $\mathbb{PQ}$-space carries a natural vector-bundle structure \cite{lee2013} over spatial positions, with a fiber of wavevectors over each position (\cref{fig:pq-bundle}). Preserving this structure prevents spatial positions and wavevectors from being mixed arbitrarily, decomposing an alignment into a spatial deformation of $\mathbb{P}$-space and a position-dependent linear transformation of $\mathbb{Q}$-space. 

Following the VoxelMorph framework \cite{balakrishnan2019}, we train a neural registration operator that maps a moving--fixed image pair to the parameters of an admissible alignment. Spatial positions and wavevectors are both expressed in the same scanner frame, meaning a rigid change of coordinates acts jointly on them. Thus, a geometry-aware registration operator should be equivariant to this joint action.

In the original VoxelMorph formulation, the registration operator uses a U-Net backbone \cite{ronneberger2015} that takes a pair of scalar-valued images as input. Our input is instead a pair of dMRI signals. We restrict our attention to the widely used single- and multi-shell settings, in which wavevectors are sampled on concentric spherical shells at fixed radii. These sampling schemes naturally lend themselves to learned feature extraction via \textit{spatio-angular convolutions}, which we use to assemble a U-Net tailored to raw dMRI.

Finally, we hope to draw the attention of the medical image analysis community to a correspondence that is well understood in equivariant deep learning \cite{cohen2019,weiler2023,bekkers2024} but has yet to find its way into dMRI analysis. A spatio-angular convolution and a 3D steerable convolution~\cite{weiler2018} are one and the same operation, and such layers therefore do not have to be designed from scratch. We use them to build what is, to our knowledge, the first equivariant registration network for raw dMRI.

\noindent\textbf{Contributions.} Our contributions in this paper are fourfold.
\begin{enumerate}
    \item We give a vector-bundle formulation of dMRI registration that makes precise what constitutes an admissible alignment.
    \item We leverage the equivalence between spatio-angular and steerable convolutions to predict admissible dMRI alignments equivariantly.
    \item We derive an alignment loss based on maximum mean discrepancy~\cite{sriperumbudur2010} that compares water-displacement distributions without explicitly reconstructing them, and a local-contrast form of it, robust to between-subject contrast differences, that reduces to a local correlation.
    \item We show, with the optimizer held fixed and against a known deformation of real tissue, that the raw signal yields more accurate alignment than a derived scalar, and we confirm the resulting method on held-out HCP subjects and a held-out clinical-grade cohort, AOMIC~\cite{snoek2021}, against classical and learned baselines.
\end{enumerate}

\section{Related Work}
\label{sec:related-work}

\paragraph{Classical dMRI registration.}
Methods for dMRI registration differ in both the representation they align and how they account for orientation. Tensor reorientation~\cite{alexander2001}, registration of Gaussian mixture fields~\cite{cheng2009}, and registration and reorientation of fiber orientation distributions~\cite{raffelt2011,raffelt2012} address these questions for derived representations. Raw-signal registration also has established precedents. Duarte-Carvajalino et al.~\cite{duarte2013} use angular interpolation to register diffusion-weighted volumes without fitting a local diffusion model, while Zhang et al.~\cite{zhang2014} incorporate signal reorientation into large-deformation diffeomorphic registration. At the tractography level, BundleWarp~\cite{chandio2026} nonlinearly aligns streamline bundles rather than voxelwise diffusion signals. Our vector-bundle formulation makes the admissible transformations of the acquisition domain explicit.

\paragraph{Data-driven dMRI registration.}
DDMReg~\cite{zhang2021} estimates spatial deformations from fractional anisotropy and tract orientation maps. MVCReg/MVVSReg~\cite{bouza2023} instead registers fiber orientation distributions using manifold-valued convolutions. FetDTIAlign~\cite{li2025fetdtialign} develops learned affine and deformable registration for fetal diffusion tensor images. These approaches learn alignment from derived representations; our network instead takes raw dMRI signals as input and predicts spatial deformation and local reorientation within the admissible class.

\paragraph{Equivariant and spatio-angular networks.}
Spatial and angular processing of dMRI is already central to networks for segmentation~\cite{muller2021,liu2023}, angular super-resolution~\cite{lyon2023}, and deconvolution~\cite{elaldi2024}. Tensor field networks~\cite{thomas2018} and 3D steerable CNNs~\cite{weiler2018} provide tools for constructing equivariant scalar and vector features. Billot et al.~\cite{billot2024} use an $\mathrm{SE}(3)$-equivariant network for global rigid motion tracking in brain MRI. Our task instead requires spatially varying deformations and reorientations of raw dMRI. We use the correspondence between spatio-angular and steerable convolutions to predict these fields equivariantly under the joint rigid action on spatial positions and wavevectors.

\section{Registration framework}
\label{sec:methodology}

Our framework separates the question of what constitutes an admissible alignment from how to estimate one. We first characterize the transformations allowed by the acquisition geometry, then construct an equivariant network to predict them and a distribution-based objective to train it.

\subsection{Admissible alignments}
\label{ssec:theory}

We denote the \textit{space of admissible alignments} by $\mathscr{A}$.

\paragraph{Geometry of dMRI.}
\label{sssec:geometry-of-dmris}

Put simply, a dMRI acquisition is a collection of 3D volumes indexed by finitely many diffusion-encoding wavevectors $\mathbf{q}_k\in\mathbb{R}^3$. The direction of each wavevector specifies the direction of an applied diffusion-sensitizing magnetic-field gradient, whose magnitude depends on the gradient strength and duration~\cite{callaghan1991}. We therefore model the acquisition continuously as a function $S\colon\mathbb{R}^3\times\mathbb{R}^3\to\mathbb{R}$. Let $\operatorname{Diff}^{+}\!(X)$ denote the group of orientation-preserving diffeomorphisms of a smooth manifold $X$~\cite[Ch.~15]{lee2013}. At face value, one is tempted to take $\mathscr{A}=\operatorname{Diff}^{+}\!(\mathbb{R}^3\times\mathbb{R}^3)$ as the space of admissible alignments between two dMRI signals $S_{\mathrm{M}}$ and $S_{\mathrm{F}}$. This choice, however, does not distinguish the roles of spatial positions and wavevectors, leading to the following two problems.
% \smallskip
\begin{enumerate}[label=P\arabic*, ref=Problem \arabic*]
    \item\label{I1} Consider $\Phi(\mathbf{p}, \mathbf{q}):=(\mathbf{p} + \mathbf{q}, \mathbf{q})$. Clearly $\Phi$ is in $\operatorname{Diff}^{+}\!(\mathbb{R}^3 \times \mathbb{R}^3)$, and yet it seems to imply a nonsensical conclusion: namely, that an input position $\mathbf{p}$ can be spread across infinitely many output positions $\mathbf{p} + \mathbf{q}$ for each $\mathbf{q} \in \mathbb{R}^3$. Moreover, it does not make physical sense to add a position vector to a wavevector.
    \item\label{I2} An important feature of a dMRI acquisition is the \textit{$b = 0$ volume}, defined as $S_0:=S(\cdot, \mathbf{0})$. It corresponds to an image acquired with \textit{no} diffusion weighting. A reasonable registration should not artificially introduce diffusion weighting into a non-diffusion-weighted volume, and yet $\Phi(\mathbf{p}, \mathbf{q}):=(\mathbf{p}, \mathbf{p} + \mathbf{q})$ does exactly that.
\end{enumerate}
% \smallskip
These concerns suggest that specifying the dMRI domain to be $\mathbb{R}^3 \times \mathbb{R}^3$ and nothing more tells an incomplete story. We propose the correct domain to be the \textit{vector bundle}~\cite[Ch.~10]{lee2013}
\begin{equation}
    \begin{tikzcd}
    \mathbb{R}^3 \times \mathbb{R}^3 \arrow[r, "\pi"] & 
    \mathbb{R}^3
    \end{tikzcd},
    \qquad \pi(\mathbf{p}, \mathbf{q}):=\mathbf{p}.
\label{eqn:vector-bundle}
\end{equation}
We refer to the vector bundle in \cref{eqn:vector-bundle} as $\mathbb{PQ}$. For the purposes of this work, the only prerequisite notions we require are (a) the \textit{fiber} over position $\mathbf{p}$ is the preimage $\pi^{-1}(\mathbf{p}) \simeq \mathbb{R}^3$, and (b) a \textit{section} of $\mathbb{PQ}$ is a map $s: \mathbb{R}^3 \to \mathbb{R}^3 \times \mathbb{R}^3$ such that $\pi \circ s = \mathrm{id}_{\mathbb{R}^3}$ \cite[Ch.~10]{lee2013}. We will write the fiber over $\mathbf{p}$ as $\mathbb{PQ}_\mathbf{p}$. Among other benefits, a vector bundle framework provides clean separation between objects that should not be mixed: $\mathbf{p}$-vectors in the base space (or scanner space) and $\mathbf{q}$-vectors in the fibers.

\paragraph{Registration of dMRI.} Having described the domain of a dMRI signal as the vector bundle $\mathbb{PQ}$, it is only natural that we define the space of admissible alignments between two dMRI signals to be the set of automorphisms of $\mathbb{PQ}$.
An \textit{automorphism} of $\mathbb{PQ}$ is a pair $(\Phi, \varphi)$ of diffeomorphisms such that the following diagram commutes:
\begin{equation}
    \begin{tikzcd}
        \mathbb{R}^3 \times \mathbb{R}^3 \arrow[r, "\Phi"] \arrow[d, "\pi"] & \mathbb{R}^3 \times \mathbb{R}^3 \arrow[d, "\pi"] \\
        \mathbb{R}^3 \arrow[r, "\varphi"] & \mathbb{R}^3
    \end{tikzcd}
\label{eqn:bundle-automorphism}
\end{equation}
Furthermore, any restriction $\Phi|_{\mathbb{PQ}_\mathbf{p}}: \mathbb{PQ}_\mathbf{p} \to \mathbb{PQ}_{\varphi(\mathbf{p})}$ to a fiber must be linear~\cite{michor2008}. We denote the space of automorphisms of $\mathbb{PQ}$ by $\operatorname{Aut}(\mathbb{PQ})$, and we write $\Phi$ instead of $(\Phi, \varphi)$ when context is clear.

\Cref{eqn:bundle-automorphism}, together with fiberwise linearity, forces $\Phi$ to have the form
\begin{equation}
    \Phi(\mathbf{p}, \mathbf{q}) = (\varphi(\mathbf{p}), T(\mathbf{p})\,\mathbf{q}),
\label{eqn:auto-structure}
\end{equation}
where $T: \mathbb{R}^3 \to \mathrm{GL}(3)$ is a smooth field of invertible linear operators given by $T(\mathbf{p}) = \Phi|_{\mathbb{PQ}_\mathbf{p}}$. We observe that \cref{eqn:auto-structure} readily resolves \ref{I1} by ensuring the first factor only depends on $\mathbf{p}$. Furthermore, observe that the $b = 0$ volume $S_0$ corresponds to the \textit{zero section} of $\mathbb{PQ}$ given by $s(\mathbf{p}):=(\mathbf{p}, \mathbf{0})$. By fiberwise linearity, the zero section is mapped to itself under an automorphism, thus settling \ref{I2}.

To summarize the situation thus far, we have reduced the space of admissible dMRI alignments from $\operatorname{Diff}^{+}\!(\mathbb{R}^3 \times \mathbb{R}^3)$ to $\operatorname{Aut}(\mathbb{PQ})$. Typically, dMRI acquisitions sample wavevectors along multiple directions at one or a few fixed magnitudes, forming concentric spherical shells in $\mathbb{Q}$-space. To align signals within corresponding shells, we further restrict $\mathscr{A}$ by requiring transformations to preserve shell radius. That is, we impose \textit{shell preservation}:
\begin{equation}
    \|T(\mathbf{p})\,\mathbf{q}\| = \|\mathbf{q}\|,\,\forall\,\mathbf{p},\mathbf{q} \in \mathbb{R}^3.
\label{eqn:shell-preservation}
\end{equation}
By orientation preservation in the fibers, \cref{eqn:shell-preservation} implies $T(\mathbf{p}) \in \mathrm{SO}(3)$ $\forall\mathbf{p} \in \mathbb{R}^3$. We denote the resulting subgroup of shell-preserving automorphisms by $\operatorname{Aut}_{\mathrm{SO}}(\mathbb{PQ})$. A direct consequence of shell preservation is that an admissible alignment cannot move signal between shells of different radii.

% At this stage, the admissible transformation set $\operatorname{Aut}_{\mathrm{SO}}(\mathbb{PQ})$ still strictly enlarges the classical diff-$\mathbf{p}$ transformation class underlying \cref{eqn:diff-p-registration-problem}. In the diff-$\mathbf{p}$ setting, one implicitly assumes that the fiber action $T(\mathbf{p})$ is determined locally by the base diffeomorphism $\varphi$, typically via the rotational part of its Jacobian $D\varphi(\mathbf{p})$. In our bundle formulation, this corresponds to a \emph{locality} axiom.
% % : the action on the fiber over $\mathbf{p}$ is allowed to depend only on the local behavior of $\varphi$ near $\mathbf{p}$, and not on the global behavior of $\varphi$ or on the particular signals $S_{\mathrm{M}},S_{\mathrm{F}}$.
% This locality requirement is precisely the additional assumption that, together with shell preservation and orientation preservation, reduces $\operatorname{Aut}_{\mathrm{SO}}(\mathbb{PQ})$ to the diff-$\mathbf{p}$ transformation class embodied in \cref{eqn:diff-p-registration-problem}.

Thus $\operatorname{Aut}_{\mathrm{SO}}(\mathbb{PQ})$ strictly enlarges the class of spatial deformations with prescribed Jacobian-based reorientation. This restricted class is recovered by imposing the additional locality axiom that $T(\mathbf p)$ is determined by the local Jacobian of $\varphi$, typically through its rotational part.

\subsection{Equivariant registration network}
\label{ssec:practice}

\paragraph{Equivariant registration.} A \textit{registration operator} $\mathsf{R}_X$ maps a moving--fixed pair of signals $(S_{\mathrm{M}},S_{\mathrm{F}})$ on a domain $X$ to an admissible alignment,
\begin{equation}
    \mathsf{R}_X: L^2(X) \times L^2(X) \to \mathscr{A} \subseteq\operatorname{Diff}^{+}\!(X).
\end{equation}
For multi-shell dMRI, we take $X=\mathbb{PQ}$ and $\mathscr{A}=\operatorname{Aut}_\mathrm{SO}(\mathbb{PQ})$.

Spatial positions and wavevectors are expressed in the same scanner frame, so a rigid change of coordinates acts jointly on them. For a roto-translation $(\mathbf{R},\mathbf{t})\in\mathrm{SE}(3)$~\cite{hall2015}, the \textit{joint action} on $\mathbb{PQ}$ is given by 
\begin{equation}
    (\mathbf{R},\mathbf{t})\cdot(\mathbf{p},\mathbf{q}):=(\mathbf{R}\mathbf{p}+\mathbf{t},\mathbf{R}\mathbf{q}).
\label{eqn:se3-joint-action}
\end{equation}
This induces the action on signals
\begin{equation*}
    [ (\mathbf{R},\mathbf{t})\cdot S ](\mathbf{p},\mathbf{q})
    :=S\bigl(\mathbf{R}^\top(\mathbf{p}-\mathbf{t}),\mathbf{R}^\top\mathbf{q}\bigr).
\end{equation*}
A registration operator should respond consistently when the same rigid change of scanner coordinates is applied to both input signals. We therefore require \textit{equivariance},
\begin{equation}
    \mathsf{R}_\mathbb{PQ}\bigl((\mathbf{R},\mathbf{t})\cdot S_{\mathrm{M}},(\mathbf{R},\mathbf{t})\cdot S_{\mathrm{F}}\bigr)
    = (\mathbf{R},\mathbf{t})\circ\mathsf{R}_\mathbb{PQ}(S_{\mathrm{M}},S_{\mathrm{F}})\circ(\mathbf{R},\mathbf{t})^{-1},
\label{eqn:equi-R_PQ}
\end{equation}
for all $S_{\mathrm{M}},S_{\mathrm{F}}\in L^2(\mathbb{PQ})$ and $(\mathbf{R},\mathbf{t})\in\mathrm{SE}(3)$. Writing $(\varphi_{(\mathbf{R},\mathbf{t})},T_{(\mathbf{R},\mathbf{t})})$ for the components predicted from $((\mathbf{R},\mathbf{t})\cdot S_{\mathrm{M}},(\mathbf{R},\mathbf{t})\cdot S_{\mathrm{F}})$ and $(\varphi,T)$ for those from the original pair, \cref{eqn:equi-R_PQ} becomes
\begin{equation}
\begin{aligned}
\varphi_{(\mathbf R,\mathbf t)}(\mathbf p)
  &= \mathbf R\,\varphi\big(\mathbf R^\top(\mathbf p-\mathbf t)\big) + \mathbf t,\\
T_{(\mathbf R,\mathbf t)}(\mathbf p)
  &= \mathbf R\,T\big(\mathbf R^\top(\mathbf p-\mathbf t)\big)\,\mathbf R^\top.
\end{aligned}
\label{eqn:equi-R_PQ-decomposed}
\end{equation}
The network construction below is designed to satisfy these two transformation laws.

\begin{figure*}[t]
 \centering
    \includegraphics[width=\textwidth]{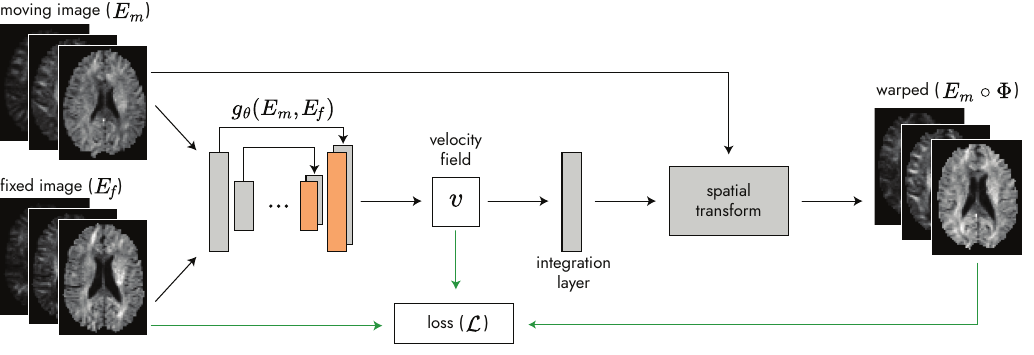}
    \caption{Proposed registration pipeline, parameterized by an $\mathrm{SE}(3)$-equivariant U-Net.}
    \label{fig:arch}
\end{figure*}

\paragraph{VoxelMorph framework.}
Following VoxelMorph \cite{balakrishnan2019}, we train a network to align moving--fixed image pairs by minimizing their dissimilarity after deformation, together with a regularization penalty. In its diffeomorphic variant \cite{dalca2019}, a U-Net predicts a stationary velocity field, which is exponentiated and used to differentiably resample the moving image \cite{jaderberg2015}. We adapt these components to raw dMRI below (\cref{fig:arch}).

% \paragraph{VoxelMorph framework.} \cref{def:registration-operator} lines up closely with the self-supervised approach to image registration introduced in \cite{balakrishnan2019}, which we refer to as the VoxelMorph framework. Abstractly, pairs $(S_{\mathrm{M}}, S_{\mathrm{F}}) \in L^2(X) \times L^2(X)$ are fed to a neural surrogate $\mathsf{R}^\Theta_X$ parameterized by $\Theta$. The goal is to best approximate the ground-truth registration operator $\mathsf{R}_X$ by training $\mathsf{R}^\Theta_X$ to minimize (ignoring regularization)
% \begin{equation}
%     \sum_{(S_{\mathrm{M}}, S_{\mathrm{F}})}\mathsf{L}\bigl(S_{\mathrm{F}}, S_{\mathrm{M}} \circ \mathsf{R}^\Theta_X(S_{\mathrm{M}}, S_{\mathrm{F}})\bigr)
% \end{equation}
% for some choice of dissimilarity measure $\mathsf{L}$. Three principal building blocks are required for this framework:
% \smallskip
% \begin{enumerate}
%     \item A neural network $\mathsf{N}^\Theta$ that estimates a stationary velocity field (SVF) \cite{arsigny2006} $v$ on $X$, typically a U-Net.
%     \item An exponentiation block $\mathsf{SS}$ that generates a diffeomorphism $\Phi:=\exp(v)$ from $v$. This is most efficiently done using \textit{scaling and squaring} \cite{arsigny2006}. We set $\mathsf{R}^\Theta_X:=\mathsf{SS}\circ \mathsf{N}^\Theta$.
%     \item A spatial transformer \cite{jaderberg2015} that deforms $S_{\mathrm{M}}$ by $\Phi$, i.e.\ a differentiable resampling module.
% \end{enumerate}
% \smallskip
% We now describe how each of these modules can be tailored to the setting of $L^2(\mathbb{PQ})$. See \cref{fig:arch} for a schematic.

\paragraph{Steerable U-Net.} As discussed in the introduction, shell-sampled dMRI naturally lends itself to spatio-angular convolutions. In spherical-harmonic coordinates, these equivariant convolutions can be implemented as 3D steerable convolutions \cite{cohen2019,weiler2023,bekkers2024}, allowing us to build on existing layers rather than design them from scratch \cite{weiler2018}.

We first compute the \textit{signal attenuation} $E=S/S_0$ and expand each shell in spherical harmonics. We model the signal as antipodally symmetric, $E(\mathbf p,\mathbf q)=E(\mathbf p,-\mathbf q)$, so only even harmonic degrees are needed \cite{elaldi2024}. Up to an even bandlimit $B$, the coefficients form a spherical tensor field $\widehat E$ transforming under
\begin{equation}
    \rho^B(\mathbf R)=D^0(\mathbf R)\oplus D^2(\mathbf R)\oplus\cdots\oplus D^B(\mathbf R),
\end{equation}
where $D^\ell$ is the degree-$\ell$ Wigner rotation matrix. For multi-shell data, we concatenate the coefficient fields along the channel axis and predict a deformation and reorientation shared across shells.

Our U-Net uses steerable convolution blocks to produce a vector field $z:\mathbb R^3\to\mathbb R^3$, transforming under $D^1$. Gaussian smoothing gives the spatial velocity $v$; its exponentiation and the reorientation below give the transformation laws in \cref{eqn:equi-R_PQ-decomposed}.

\paragraph{Exponentiation and reorientation.} The velocity $v$, of finite Sobolev norm \cite{younes2010}, is
exponentiated by scaling and squaring \cite{arsigny2006,dalca2019}: the time-$1$ flow $\varphi=\exp(v)$ of
$\tfrac{d}{dt}\phi_t=v\circ\phi_t$, $\phi_0=\mathrm{id}$, is approximated by $K$ squaring steps starting from
$\mathbf p\mapsto\mathbf p+2^{-K}v(\mathbf p)$, each implemented by a differentiable spatial transformer. The flow is
natural under diffeomorphisms of the domain \cite[Ch.~9]{lee2013},
$\exp\bigl((\mathbf R,\mathbf t)\cdot v\bigr)=(\mathbf R,\mathbf t)\circ\exp(v)\circ(\mathbf R,\mathbf t)^{-1}$, so
$\varphi$ satisfies \cref{eqn:equi-R_PQ-decomposed}, and for a smooth $v$ it is a diffeomorphism with
$\det D\varphi>0$. We take $T(\mathbf p)=\operatorname{polar}\bigl(D\varphi(\mathbf p)\bigr)\in\mathrm{SO}(3)$, the
rotation factor of the Jacobian (finite-strain reorientation \cite{alexander2001}), so that $T$ is determined by the
local Jacobian of $\varphi$ as in \cref{ssec:theory}. Under a rigid change of coordinates $D\varphi$ is conjugated by
$\mathbf R$, and the polar factor commutes with this conjugation, so $T$ satisfies \cref{eqn:equi-R_PQ-decomposed}.

\paragraph{Deformation.} On a single shell, $(\varphi,T)$ acts on the moving coefficient field by a spatial warp and a
pointwise rotation of the coefficients,
\begin{equation}
    \widehat{E}_{\mathrm{M}}^\Phi(\mathbf p)
    := \rho^B\bigl(T(\mathbf p)\bigr)\,\widehat{E}_{\mathrm{M}}\bigl(\varphi^{-1}(\mathbf p)\bigr),
    \qquad
    \rho^B(T) = D^0(T)\oplus D^2(T)\oplus\cdots\oplus D^B(T),
\end{equation}
implemented by a spatial transformer on the coefficient field followed by multiplication with the Wigner $D$-matrices
of $T(\mathbf p)$.

\subsection{A Novel Loss for dMRI Registration}
\label{ssec:loss}

Voxelwise squared differences of diffusion-weighted volumes and similarities of derived representations (e.g.\ FA,
fODF) neglect the probabilistic interpretation of the diffusion signal. We instead derive a loss that targets the
alignment of ensemble average propagators (EAPs) and remains a weighted $L^2$ loss on the signal attenuations. For each
$\mathbf p$, the attenuation $\mathbf q\mapsto E(\mathbf p,\mathbf q)$ is (up to normalization) the characteristic
function of a probability density on displacement space, the EAP
$P_{\mathbf p}(\mathbf r)=\int_{\mathbb{R}^3} e^{2\pi i\,\mathbf q^\top \mathbf r}E(\mathbf p,\mathbf q)\,d\mathbf q$.

Given fixed and moving dMRIs $S_{\mathrm{F}}$ and $S_{\mathrm{M}}$, with corresponding attenuations
$E_{\mathrm{F}}$ and $E_{\mathrm{M}}$, and a deformation $\Phi$ acting on $\mathbb{PQ}$,
the ideal registration objective is to minimize a discrepancy between the EAPs
$P^{\mathrm{F}}_{\mathbf p}$ and $P^{\mathrm{M}}_{\mathbf p}$ for all $\mathbf p$. To this end, we
consider the maximum mean discrepancy (MMD) between EAPs at each voxel
$\mathbf p$ with respect to a Gaussian kernel
$k(\mathbf r,\mathbf r') = \exp\bigl(-\|\mathbf r-\mathbf r'\|^2/(2\sigma^2)\bigr)$.
By the theory of kernel mean embeddings of probability measures
(cf.\ \cite{sriperumbudur2010}), the squared MMD admits an equivalent
expression in terms of characteristic functions (shown here in discrete form):
% \begin{equation}
% \label{MMDloss}
% \operatorname{MMD}^2\!\left(P^{\mathrm{F}}_{\mathbf p}, P^{\mathrm{M}}_{\mathbf p}\right)
% \propto
% \int_{\mathbb{R}^3}
%     \bigl| E_{\mathrm{F}}(\mathbf p,\mathbf q)
%            - [E_{\mathrm{M}} \circ \Phi](\mathbf p,\mathbf q)\bigr|^2\,
%     w(\mathbf q)\,d\mathbf q.
% \end{equation}
% where the weight
% $w(\mathbf q) = \|\mathbf q\|^2 \exp\bigl(-\|\mathbf q\|^2/(2\sigma^2)\bigr)$
% depends only on the radius $\|\mathbf q\|$.

% Discretizing over voxels $\mathbf p$ and sampled diffusion wavevectors
% $\mathbf q \in \mathcal{Q}$, we obtain the following loss on the attenuations:
\begin{equation}
\label{eqn:mmd-loss}
%\begin{aligned}
\mathsf{L}_{\text{MMD}}(E_{\mathrm{F}}, E_{\mathrm{M}} \circ \Phi)
   % & 
    =
    C \sum_{\mathbf p}\sum_{\mathbf q \in \mathbb{Q}}
    \Bigl| E_{\mathrm{F}}(\mathbf p,\mathbf q)
           - [E_{\mathrm{M}} \circ \Phi](\mathbf p,\mathbf q)\Bigr|^2 %\\
   % &
    %\hspace{3.0em}
  %  \times
    \exp\!\Bigl(-\tfrac{\|\mathbf q\|^2}{2\sigma^2}\Bigr)\,
    \|\mathbf q\|^2,
%\end{aligned}
\end{equation}
where $C = (2\pi\sigma^2)^{-3/2}$ and $\sigma > 0$ is the kernel bandwidth. The inner sum runs over the sampled shells
and directions, and $E_{\mathrm{M}}\circ\Phi$ is synthesized from the deformed coefficient field at the sampled
wavevectors; the derivation depends only on the relation between $E(\mathbf p,\cdot)$ and $P_{\mathbf p}$, not on the
basis used to store $E$. The weight is constant on each shell, so on sampled shells \cref{eqn:mmd-loss} is a squared
attenuation difference with one weight per shell, set by $\sigma$. Although it resembles a weighted sum of squared
differences, its weights must be the Fourier transform of a positive definite kernel, with finite total integral.

\paragraph{Local contrast.} \Cref{eqn:mmd-loss} presumes that both subjects share the same contrast. Between subjects
and scanners, however, $T_2$ weighting, partial volume and coil sensitivity change the attenuations by spatially varying
gains and offsets that no deformation can explain, so we let the comparison absorb a local affine change of contrast. For a sampled wavevector $\mathbf q$ and a window
$W(\mathbf p)$ of voxels centered at $\mathbf p$, let $\mathbf e_{\mathrm F},\mathbf e_{\mathrm M}\in\mathbb R^{|W|}$
collect $E_{\mathrm F}(\cdot,\mathbf q)$ and $[E_{\mathrm M}\circ\Phi](\cdot,\mathbf q)$ over the window. Minimizing the
squared discrepancy of \cref{eqn:mmd-loss} over a gain $\alpha$ and an offset $\beta$, relative to the local
variation of the fixed attenuation, has the closed form
\begin{equation}
\label{eqn:lc}
    \min_{\alpha,\beta\in\mathbb R}
    \frac{\|\mathbf e_{\mathrm F}-\alpha\,\mathbf e_{\mathrm M}-\beta\mathbf 1\|^2}
         {\|\mathbf e_{\mathrm F}-\bar e_{\mathrm F}\mathbf 1\|^2}
    = 1-\rho^2(\mathbf p,\mathbf q),
\end{equation}
where $\rho$ is the correlation of $\mathbf e_{\mathrm F}$ and $\mathbf e_{\mathrm M}$ over the window. The
normalization does not depend on $\Phi$, so the resulting loss
\begin{equation}
\label{eqn:lc-loss}
    \mathsf{L}_{\text{LC}}(E_{\mathrm{F}}, E_{\mathrm{M}} \circ \Phi)
    =
    \frac{1}{|\mathbb Q|}\sum_{\mathbf p}\sum_{\mathbf q \in \mathbb{Q}} \bigl(1-\rho^2(\mathbf p,\mathbf q)\bigr)
\end{equation}
uses the locally normalized residual after the best affine contrast correction, with uniform wavevector weights.
Fixing $\alpha=1$, $\beta=0$, dropping the local normalization, and restoring the weighting and overall scaling of
\cref{eqn:mmd-loss} recovers that expression. As $1-\rho^2$ is invariant to the scale of each shell, we weight
the sampled wavevectors uniformly. \Cref{eqn:lc-loss} is the local correlation of SyN \cite{avants2008} applied to each
sampled wavevector. We refer to \cref{eqn:mmd-loss} as the \emph{global} MMD and to
\cref{eqn:lc-loss} as the MMD \emph{with local contrast}, which is our default (\cref{ssec:ablations} compares them).

Our total loss augments the similarity $\mathsf{L}_{\text{sim}}\in\{\mathsf{L}_{\text{MMD}},\mathsf{L}_{\text{LC}}\}$
with a Sobolev penalty on the velocity,
\begin{equation}
    \mathsf{L}_{\text{reg}}(v)
    =
    \lambda_v \sum_{\mathbf p} \|\nabla v(\mathbf p)\|^2_F,
\end{equation}
and we minimize
$
    \mathsf{L}_{\text{total}}
    =
    \mathsf{L}_{\text{sim}}(E_{\mathrm{F}}, E_{\mathrm{M}} \circ \Phi)
    +
    \mathsf{L}_{\text{reg}}(v).
$
The velocity is Gaussian-smoothed before exponentiation, $v=G_s*z$, where $G_s$ is a Gaussian kernel of standard
deviation $s$ and $z$ is the predicted (or optimized) field, as in SyN's update smoothing. Weights are chosen on validation pairs (\cref{ssec:model-training-and-setup}).

\section{Experiments}
\label{sec:experiments}

We ask three questions. (i)~Does the proposed pipeline register held-out data accurately, at research and at clinical
grade? (ii)~What does the steerable network contribute relative to a conventional U-Net, and how much of the final
accuracy comes from refinement? (iii)~Which components of the pipeline matter: the raw-signal input, the objective and
the reorientation?

\subsection{Setup}
\label{ssec:model-training-and-setup}

\paragraph{Data.}
We split $50$ minimally preprocessed HCP subjects \cite{vanessen2012} ($b=1000,2000,3000$\,s/mm$^2$, $90$ directions
each) into $30/10/10$ training/validation/test subjects and train on $450$ randomly selected ordered training pairs. All
design choices were made on the validation subjects. The test subjects were scored once, at native resolution
($1.25$\,mm), with all methods and settings frozen beforehand; each test subject is fixed once and moving once ($10$
pairs). As a clinical-grade cohort we use AOMIC-PIOP1 \cite{snoek2021} (Philips 3T, a single $b=1000$\,s/mm$^2$ shell
with $32$ directions, about $2$\,mm voxels): $40$ healthy adults from a different scanner, site and population, never
used for training. AOMIC images are rigidly aligned to the HCP grid and registered at $2.5$\,mm; each subject is fixed
once and moving once ($40$ pairs). Because its direction scheme covers the sphere non-uniformly, spherical-harmonic fits
on AOMIC use Laplace--Beltrami regularization \cite{descoteaux2007} ($\lambda=0.006$). HCP data are used under the
WU-Minn HCP Open Access Data Use Terms; AOMIC-PIOP1 is distributed under CC0.

\paragraph{Method.}
After a raw-signal affine pre-alignment, the steerable U-Net of \cref{sec:methodology} (three resolution levels, width
$32$, $2.7\times10^5$ parameters) predicts the velocity from per-shell spherical-harmonic coefficients (degree $8$; $6$ for
single-shell data) of both attenuations; the moving signal is reoriented by finite strain. The objective is the MMD with local contrast
(\cref{eqn:lc-loss}, $9^3$-voxel windows) with a Gaussian-smoothed velocity ($s=\sqrt3$ voxels) and a Sobolev penalty.
Networks are trained at $2.5$\,mm for $30$k steps (Adam, learning rate $10^{-3}$, gradient clipping). At test time the
network's warp is refined coarse-to-fine with the same objective ($100$ iterations per pyramid level, at the resolution
of the evaluation). For AOMIC, the same recipe is trained on HCP reduced to its $b=1000$ shell and $30$
directions.

\paragraph{Baselines and metrics.}
We compare with ANTs SyN on FA \cite{avants2008}, \texttt{mrregister} on fODFs \cite{raffelt2011,tournier2019}, and
DDMReg \cite{zhang2021} with its released weights at its own resolution ($1.25$\,mm); on AOMIC also SynthMorph
\cite{hoffmann2021} on FA with its released weights. All warps are converted to a common convention and scored with the
same code. Our primary metric is Dice over $22$ FreeSurfer structures \cite{fischl2002} (cerebral and cerebellar white
matter, corpus callosum, deep grey nuclei, hippocampus, amygdala, ventricles, brainstem), labels that no method uses as
input. We also report Dice over $72$ TractSeg bundles \cite{wasserthal2018} and, on HCP, tract distance \cite{zhang2021}
(fixed-subject streamlines mapped by the warp, mean closest-point distance to the moving subject's streamlines). DDMReg's
inputs (tract orientation maps) come from the same TractSeg model as the bundles, so the bundle readout favors it. We
report means and, for each baseline, the mean paired difference and the number of pairs in which ours is better.

\subsection{Held-Out Evaluation}
\label{ssec:native-hcp}

\begin{table}[t]
    \centering
    \small
    \setlength{\tabcolsep}{4pt}
    \caption{HCP test set at native resolution ($10$ pairs, scored once). Means; time per pair (GPU: one NVIDIA B200). The last column gives
    the paired structure-Dice difference of our method against each baseline (pairs in which ours is better).
    $^\dagger$Favors DDMReg (see text).}
    \label{tab:native_hcp}
    \resizebox{\textwidth}{!}{%
    \begin{tabular}{@{}lccccc@{}}
        \toprule
        Method & Structure Dice $\uparrow$ & Bundle Dice$^\dagger$ $\uparrow$ & Tract dist.\ (mm) $\downarrow$ & Time & Ours $-$ method \\
        \midrule
        DDMReg & $0.734$ & $0.793$ & $3.830$ & $6$\,s (GPU) & $+0.047$ ($10$) \\
        \texttt{mrregister} & $0.753$ & $0.791$ & $3.718$ & $16$\,min (4 CPU) & $+0.028$ ($10$) \\
        ANTs SyN, FA & $0.758$ & $0.811$ & $\mathbf{3.695}$ & $16$\,min (4 CPU) & $+0.023$ ($10$) \\
        Ours & $\mathbf{0.781}$ & $\mathbf{0.832}$ & $3.698$ & $3.2$\,min (GPU) & -- \\
        \bottomrule
    \end{tabular}}
\end{table}

\begin{table}[t]
    \centering
    \small
    \setlength{\tabcolsep}{4pt}
    \caption{HCP test set: orientation and microstructure agreement in the fixed white matter, and warp regularity
    within the brain ($10$ pairs, means). Every warp is reoriented by finite strain from its own Jacobian. ACC: fODF
    angular correlation; peak: angle between principal fODF peaks; OVL: tensor overlap; FA, MD (relative) and RISH
    (relative, per shell and degree) errors are rotation invariant. $\det D\varphi$: $1$st--$99$th percentiles.}
    \label{tab:hcp_micro}
    \resizebox{\textwidth}{!}{%
    \begin{tabular}{@{}lcccccccc@{}}
        \toprule
        Method & ACC $\uparrow$ & Peak ($^\circ$) $\downarrow$ & OVL $\uparrow$ & FA err.\ $\downarrow$ & MD err.\ $\downarrow$ & RISH err.\ $\downarrow$ & $\det D\varphi$ & Folds \\
        \midrule
        DDMReg & $0.439$ & $34.2$ & $0.623$ & $0.119$ & $0.170$ & $0.321$ & $0.32$--$2.69$ & $0$ \\
        \texttt{mrregister} & $0.484$ & $32.1$ & $0.649$ & $0.125$ & $0.169$ & $0.328$ & $0.62$--$1.55$ & $0$ \\
        ANTs SyN, FA & $0.501$ & $31.3$ & $0.664$ & $0.106$ & $0.149$ & $0.304$ & $0.51$--$1.79$ & $0$ \\
        Ours & $\mathbf{0.611}$ & $\mathbf{25.9}$ & $\mathbf{0.729}$ & $\mathbf{0.096}$ & $\mathbf{0.115}$ & $\mathbf{0.287}$ & $0.27$--$2.89$ & $0$ \\
        \bottomrule
    \end{tabular}}
\end{table}

\begin{figure}[t]
    \centering
    \includegraphics[width=\textwidth]{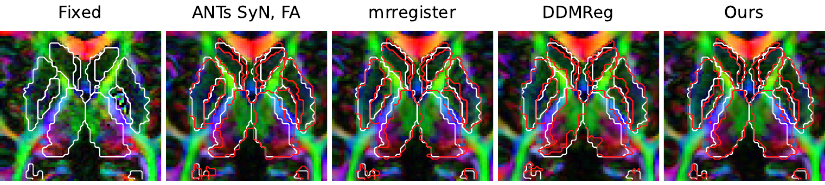}
    \caption{HCP test pair with the median structure-Dice gain of our method over ANTs SyN. Axial crops of the moving
    subject warped by each method, colored by principal diffusion direction (red: left--right, green:
    anterior--posterior, blue: inferior--superior) and scaled by FA. White: fixed FreeSurfer outlines (caudate, putamen,
    pallidum, thalamus, lateral ventricles); red: the moving subject's outlines after warping.}
    \label{fig:hcp-qualitative}
\end{figure}

\begin{table}[t]
    \centering
    \small
    \setlength{\tabcolsep}{4pt}
    \caption{AOMIC-PIOP1 ($40$ held-out pairs, single shell, inference only). Means; time per pair ($^{*}$four CPU
    cores, otherwise one GPU). The last column gives the paired structure-Dice difference of our method against each
    baseline (pairs in which ours is better, of $40$). $^\dagger$Favors DDMReg.}
    \label{tab:aomic}
    \begin{tabular}{@{}lccccc@{}}
        \toprule
        Method & Structure Dice $\uparrow$ & Hippo.\ Dice $\uparrow$ & Bundle Dice$^\dagger$ $\uparrow$ & s/pair & Ours $-$ method \\
        \midrule
        Rigid only & $0.666$ & $0.644$ & $0.668$ & -- & -- \\
        ANTs SyN, FA & $0.757$ & $0.749$ & $0.801$ & $135^{*}$ & $+0.019$ ($37$) \\
        SynthMorph, FA & $0.761$ & $\mathbf{0.767}$ & $0.773$ & $237^{*}$ & $+0.015$ ($31$) \\
        DDMReg & $0.761$ & $0.743$ & $\mathbf{0.815}$ & $7$ & $+0.015$ ($35$) \\
        \texttt{mrregister} & $0.768$ & $0.747$ & $0.776$ & $127^{*}$ & $+0.008$ ($29$) \\
        Ours & $\mathbf{0.776}$ & $0.765$ & $0.801$ & $14$ & -- \\
        \bottomrule
    \end{tabular}
\end{table}

On the HCP test set (\cref{tab:native_hcp,fig:hcp-qualitative}), our method improves structure Dice over every
baseline on all $10$ pairs and bundle Dice on $9$--$10$ pairs; DDMReg, the best method on AOMIC bundles, is last
on HCP structure Dice. Our method's tract distance is lower than DDMReg's ($9$ of $10$
pairs), but we find no clear tract-distance advantage over ANTs SyN ($+0.003$\,mm, $5$ of $10$ pairs) or
\texttt{mrregister} ($-0.020$\,mm). Our warps also give better orientation and microstructure agreement than every
baseline on all $10$ pairs and all six readouts of \cref{tab:hcp_micro}. The orientation readouts (ACC, peak, OVL)
partly reward what our objective optimizes, since a dense warp can rotate locally to match orientations; the FA, MD
and RISH errors are unaffected by reorientation. No folding was observed on HCP, but our warps change local volume more
than the classical ones (Jacobian determinant $1$st--$99$th percentiles of $0.27$--$2.89$ versus $0.51$--$1.79$ for ANTs SyN). On AOMIC (\cref{tab:aomic}), the network trained on
HCP alone improves structure Dice over every baseline, by $0.008$--$0.019$. Two exceptions qualify this: SynthMorph,
trained on brain anatomy, is the most accurate method on the hippocampus, and DDMReg is the most accurate on the bundle
readout. Folding on AOMIC is negligible (at most $0.012\%$ of
voxels, in $2$ of $40$ pairs for ours and $3$ for SynthMorph).

\subsection{What the Steerable Network Contributes}
\label{ssec:equivariance}

\begin{table}[t]
    \centering
    \small
    \setlength{\tabcolsep}{4pt}
    \caption{Steerable versus plain U-Net (HCP validation, $10$ pairs; Dice / tract distance in mm). All networks are
    trained identically (data, pre-alignment, objective, velocity smoothing, penalty, $30$k steps). Rotated: inputs and
    pre-alignment rotated by each of the three $180^\circ$ rotations about the grid axes, network only, warps mapped
    back. Refinement: coarse-to-fine with the same objective, at $2.5$\,mm or at native $1.25$\,mm resolution.}
    \label{tab:equivariance}
    \resizebox{\textwidth}{!}{%
    \begin{tabular}{@{}lccccc@{}}
        \toprule
        & & \multicolumn{2}{c}{Network only ($2.5$\,mm)} & \multicolumn{2}{c}{Network + refinement} \\
        \cmidrule(lr){3-4}\cmidrule(l){5-6}
        Network & Parameters & Canonical & Rotated (Dice) & $2.5$\,mm & $1.25$\,mm \\
        \midrule
        Plain U-Net & $471$k & $0.793$ / $3.656$ & $0.782$ & $\mathbf{0.812}$ / $\mathbf{3.573}$ & $\mathbf{0.834}$ / $\mathbf{3.554}$ \\
        Steerable, width 16 & $70$k & $0.790$ / $3.662$ & $0.790$ & $0.809$ / $3.587$ & -- \\
        Steerable, width 32 & $273$k & $\mathbf{0.794}$ / $\mathbf{3.648}$ & $\mathbf{0.794}$ & $0.810$ / $3.580$ & $0.833$ / $3.563$ \\
        \bottomrule
    \end{tabular}}
\end{table}

HCP and AOMIC images are delivered in a canonical frame, so this comparison tests robustness more than accuracy
(\cref{tab:equivariance}). With $58\%$ of the plain U-Net's parameters, the steerable network matches it on canonical
inputs and is unaffected by the tested rotations, whereas the plain network loses $0.011$ Dice; at width $16$ ($15\%$
of the parameters) it remains within $0.003$. The comparison does not show an accuracy gain from steerability on
canonically oriented data. Refinement adds $0.016$--$0.019$ Dice to each network at
$2.5$\,mm, and after refinement the plain and steerable pipelines are within $0.002$ Dice and $0.009$\,mm, the plain one
marginally ahead; on AOMIC both give structure Dice $0.776$.

\subsection{Component Ablations}
\label{ssec:ablations}

\begin{table}[t]
    \centering
    \small
    \setlength{\tabcolsep}{4pt}
    \caption{Ablations on HCP validation pairs ($2.5$\,mm, $10$ pairs; the objective optimized per pair, without a
    network). Each block changes one
    component of the final configuration (marked $\star$).}
    \label{tab:ablations}
    \begin{tabular}{@{}llcc@{}}
        \toprule
        Component & Variant & Bundle Dice $\uparrow$ & Tract dist.\ (mm) $\downarrow$ \\
        \midrule
        Input & FA, same solver and objective & $0.793$ & $3.689$ \\
              & raw dMRI$^\star$ & $\mathbf{0.816}$ & $\mathbf{3.567}$ \\
              & \textit{ANTs SyN on FA (reference)} & \textit{0.790} & \textit{3.610} \\
        \midrule
        Similarity & squared difference, uniform shell weights & $0.804$ & $3.623$ \\
                   & MMD with local contrast (\cref{eqn:lc-loss}) & $0.813$ & $3.603$ \\
                   & MMD with local contrast + velocity smoothing$^\star$ & $\mathbf{0.816}$ & $\mathbf{3.567}$ \\
        \midrule
        Reorientation & none & $0.811$ & $3.614$ \\
                      & finite strain$^\star$ & $\mathbf{0.816}$ & $\mathbf{3.567}$ \\
        \bottomrule
    \end{tabular}
\end{table}

\Cref{tab:ablations} isolates the components. Under the same solver and tuning protocol (best smoothness weight for
each input), registering the raw signal outperforms registering FA. On five real HCP subjects, each deformed by a known smooth
warp, the raw signal recovers the deformation with an endpoint error of $0.15$ voxels, against $0.38$
for FA through the same solver and $0.59$ for ANTs SyN on FA. The local contrast correction improves on the
uniformly weighted squared difference at the same regularization weight; velocity smoothing, a change of the solver
rather than of the objective, adds to it (it was not tested with the squared difference). Finite-strain reorientation
improves both readouts over no reorientation on all $10$ pairs.

\subsection{End-to-End Runtime}
\label{ssec:runtime}

\Cref{tab:runtime} reports a separate preparation-inclusive timing benchmark on $10$ HCP subjects and their $90$
ordered pairs. Preparation times are averaged over subjects and registration times over pairs, using four CPU cores
for multithreaded preparation and one NVIDIA B200 for GPU operations. These measurements concern the original
feed-forward pipeline, rather than the pipeline with iterative refinement evaluated above.

\begin{table}[!t]
    \centering
    \small
    \caption{Preparation-inclusive runtime in seconds in the separate $90$-pair HCP benchmark. Total is effective
    preparation plus registration under the concurrency assumptions described in the text.}
    \label{tab:runtime}
    \begin{tabular}{@{}lrrr@{}}
        \toprule
        Method & Effective preparation & Registration & Total \\
        \midrule
        Ours & $0.61$ & $4.76$ & $5.37$ \\
        DDMReg & $610.40$ & $1.09$ & $611.49$ \\
        \texttt{mrregister} & $469.32$ & $852.43$ & $1321.75$ \\
        \bottomrule
    \end{tabular}
\end{table}

Effective preparation assumes concurrent processing of the moving and fixed images, with dependent stages sequential.
For DDMReg, the tensor-to-FA and fODF-to-peaks-to-tract-orientation-map branches are also assumed concurrent. Our
preparation consists of attenuation computation and spherical-harmonic fitting. DDMReg has faster registration alone,
but our method has the lowest preparation-inclusive total in this benchmark; caching derived inputs changes this
comparison.

\paragraph{Limitations.}
Our framework currently restricts registration to corresponding shells with matching $b$-values; it does not support
mapping between shells of different radii.

\section{Conclusion}
\label{sec:conclusion}
In this paper, we have introduced a novel framework for registering raw dMRI data that more directly leverages orientational information. A central contribution is our construction of the registration network through the equivalence between equivariant spatio-angular convolutions and 3D steerable convolutions on spherical-harmonic coefficient fields. Controlled comparisons show that the raw signal, which contains the orientation information that derived scalars discard, yields more accurate alignment than fractional anisotropy when the optimizer is held fixed. On held-out HCP subjects and on a held-out clinical-grade cohort from a different scanner, the method improves anatomical overlap over classical and learned derived-representation baselines, and its steerable network matches a conventional U-Net with fewer parameters while being unaffected by the tested grid rotations of its inputs.

\clearpage
\begin{ack}
This research was supported in part by NIH grant RO1 NS121099 to Baba C. Vemuri and by a McKnight Doctoral
Fellowship through the Florida Education Fund to Gianfranco Cort\'{e}s.

The authors declare no competing interests.
\end{ack}

{
  \small
  \bibliographystyle{plainnat}
  \bibliography{references}
}

% \clearpage
% \input{checklist.tex}

\end{document}